\input{aipcheck}

\documentclass[
   ,final            
  ]
  {aipproc}

\layoutstyle{8x11double}

\newcommand\fverb{\setbox\fverbbox=\hbox\bgroup\verb}
\newcommand\fverbdo{\egroup\medskip\noindent%
			\fbox{\unhbox\fverbbox}\ }
\newcommand\fverbit{\egroup\item[\fbox{\unhbox\fverbbox}]}
\newbox\fverbbox

\begin{document}

\title{Black Holes at the LHC: Progress since 2002}


\author{Seong Chan Park}{
  address={
FRDP, Department of Physics and Astronomy,
Seoul National University,
Seoul, Korea}}

\begin{abstract}
  We review the recent noticeable progresses  in black hole physics
  focusing on the up-coming super-collider, the LHC.
  We discuss the classical formation of black holes by particle collision, the greybody factors
  for higher dimensional rotating black holes, the deep implications of black hole physics to the `energy-distance' relation,
  the security issues of the LHC associated with black hole formation and the newly developed Monte-Carlo generators for black hole events.
\end{abstract}

\maketitle

\section{Introduction}

\begin{figure}
    \includegraphics[width=.35\textheight]{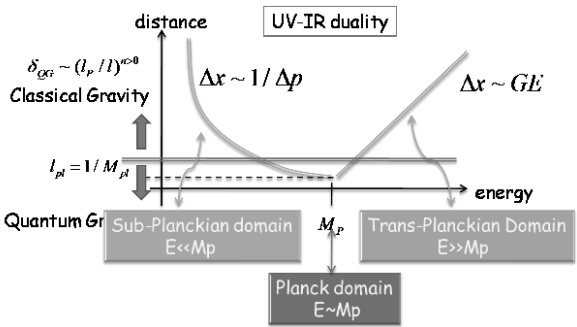}
  \caption{Energy-Distance relation is given. The horizontal axis is for the amount of energy
  we can have and the vertical axis is for the smallest distance scale we can probe with the
  given energy.}
\label{fig1_SUSY08_park}
\end{figure}
\begin{figure}
  \includegraphics[width=.35\textheight]{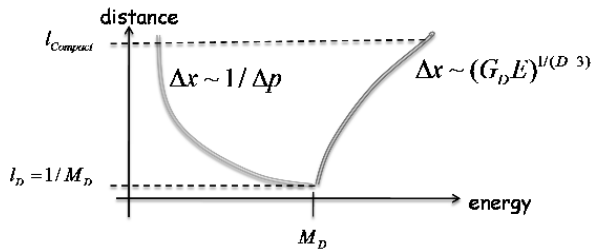}
  \caption{The same figure as \ref{fig1_SUSY08_park} when the extra dimension is taken into account.
  If there are extra dimensions the figure should be modified. Firstly, below the compactification radius, $l_{compact}$, the theory becomes higher dimensional instead of four dimensional  so that $M_{\rm Planck}$ should be understood as $M_D$ the D-dimensional true Planck scale. The Schwarzschild radius varies as $R_S\sim E^{1/(D-3)}$ in D-dimensional spacetime.}
\label{fig2_SUSY08_park}
\end{figure}

As the LHC begins to operate, we are facing various optimistic and some pessimistic
expectations about the LHC's  ability to find something interesting. One of the most spectacular
possibilities we have is that the LHC will actually be a black hole factory and produce
copious numbers of black holes. In this talk, let us
summarize what we have learned the last five years after the last plenary
talk by Greg Lansberg in the SUSY conference in 2002  ~\cite{Landsberg}.

As long as particles collide in sub-Planckian energy domain where the center of mass (CM)
energy is lower than the Planck scale, the conventional low-energy effective field theory
description is known to be valid. However, once the CM energy exceeds the Planck scale the
collision inevitably start to produce black holes and this ultimately dominates over all other
interaction processes ~\cite{'tHooft}. The corresponding computations are technically formidable
when the CM energy is just around the Planck energy ($\sqrt{s}\sim M_{\rm Planck}$) since the
physics can be described only with the underlying quantum gravity theory, such as string
theory, but once we access the transplanckian domain ($\sqrt{s}\gg M_{\rm Planck}$), a semiclassical
description of the scattering process becomes valid since the quantum gravity effects are suppressed by an order of $(l_{\rm Planck}/R_S)^{P>0}$ with an positive power \footnote{The exact power ($P$) depends on the details of the case.} where $R_S$ is the Schwarzschild radius
corresponding to the CM energy which is now the relevant length scale for the process and is much larger than the Planck length ($l_{\rm Planck}= 1/M_{\rm Planck}$). The whole picture of the high energy scattering with respect to the CM energy from the subplanckian to the transplanckian
 domain can be beautifully summarized in the energy-distance relation depicted in Fig.\ref{fig1_SUSY08_park}. Here the horizontal axis is for the
 available amount of energy for the collision and the vertical axis is for the smallest distance we can probe by the scattering process.
 In the subplanckian domain ($\sqrt{s} \ll M_{\rm Planck}$) that, the larger the CM energy the smaller distance scales can be probed
 following the uncertainty principle. Once the collision energy becomes much larger than the Planck energy, black holes form and the distance scale
 below the Schwarzschild radius is hidden below the horizon. One should note that the Schwarzschild radius increases with the energy ($R_S\sim G E$).  Once we access the length scale below the compactification scale($l_{\rm compact}$), the corresponding physics
becomes higher dimensional where the gravity scale should be replaced by the true Planck scale $M_D$
in $D$ dimensional spacetime. The size of $M_D$ is severely constrained by observational data and the lowest allowed value is lowered to the
vicinity of a TeV ~\cite{PDB2008} which is realized in large volume~\cite{ADD,AADD} or highly warped ~\cite{RS} extra dimensional models.
In TeV gravity scenarios, the LHC can be a black hole factory ~\cite{Giddings:2001bu,Dimopoulos:2001hw}.


Since the SUSY02 conference where G. Landsberg gave a plenary talk to review the status of understanding at the time around 2002,
there has been big progress in understanding black hole formation by high energy collision and the decay through Hawking radiation.

\begin{figure}
    \includegraphics[width=.35\textheight]{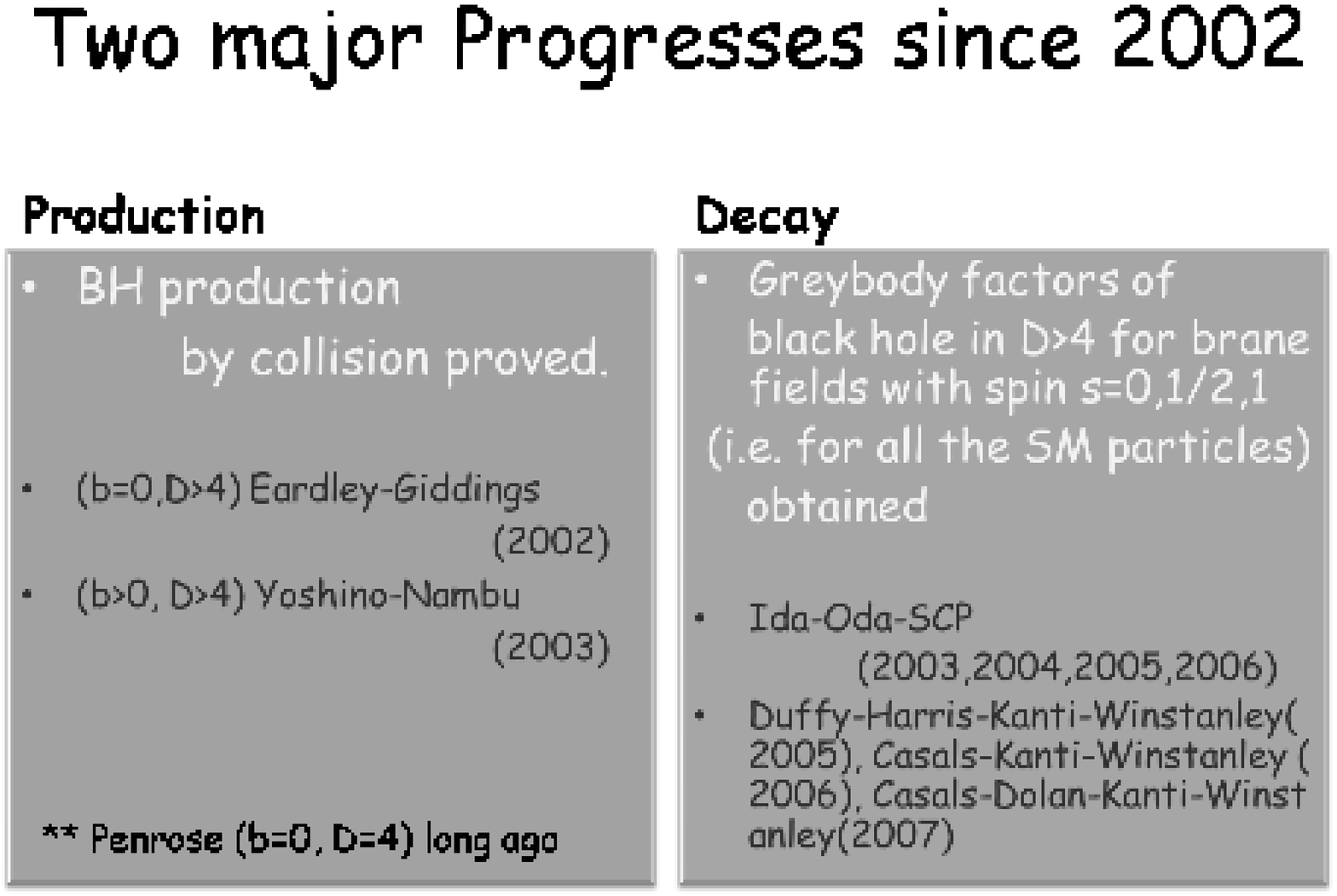}\\
  \caption{Since 2002, we have observed two main progresses both in `production' and
  `decay' of black holes in particle collision. The Hoop-conjecture is precisely confirmed
  in the case of two particle collisions and the Hawking radiation of higher dimensional
  black hole is described at the reliable level by calculation of greybody factors.}
\label{summary}
\end{figure}

\section{Progress-1: Formation}
\begin{figure}
    \includegraphics[width=.25\textheight]{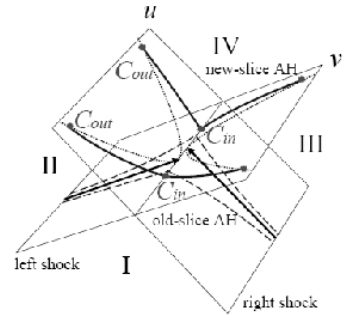}\\
  \caption{Eardley and Giddings (2002) and Yoshino and Nambu(2003) clearly demonstrated
  the formation of closed trapped surface in particle collisions. The area theorem
  tells us that classically the horizon area of the ultimate black hole must be
  greater than the original closed trapped surface. Thus black hole forms. The Figure originally comes from Yoshino and Rychkonv (2005)~\cite{formation3}. }
\label{yoshino}
\end{figure}

The classical formation of a black hole by particle collision has been explicitly shown by Eardley and Giddings (2002)~\cite{formation1}, Yoshino and Nambu (2003) ~\cite{formation2} and Yoshino and Rychkov (2005) ~\cite{formation3} for non-zero impact parameter and general dimension. They confirmed the
expectation by the long standing `Hoop conjecture ' ~\cite{hoop} which provides the clear criterion to make a black hole by compactifying energy into a small space. The Hoop conjecture states: "An imploding object forms a black hole when, and only when, a circular hoop with a specific critical circumference can be placed around the object and rotated. The critical circumference is given by 2 times $\pi$ times the Schwarzschild radius corresponding to the object's mass". The rigorous proof of Hoop conjecture is still unavailable. What has been shown is clearly depicted in Fig. ~\ref{yoshino}. In their construction, two Aichelburg-Sexl solutions, which represent two colliding particles, actually form a closed trapped surface, once the impact parameter is small enough, entirely in the well-understood geometry. The area theorem \footnote{The area theorem states that the horizon area of the final black hole must be larger than that of the initial closed trapped surface.} of general relativity then guarantees that this corresponds to formation of a black hole. The upper bound for the impact parameter is shown to be well approximated by the Schwarzschild radius of the corresponding CM energy.

\begin{figure}[tb]
    \includegraphics[width=.35\textheight]{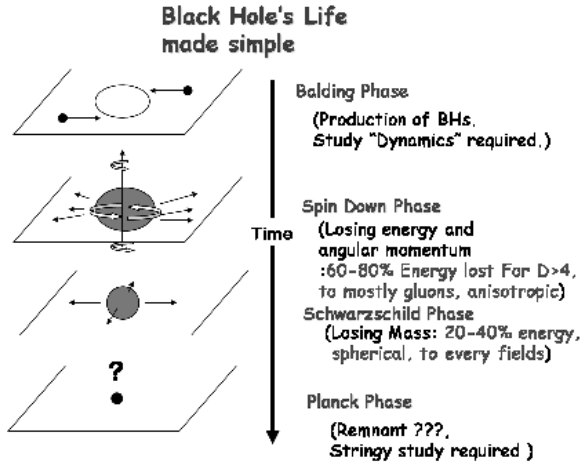}
  \caption{A Black hole evolves via four steps-Balding, spindown, Schwarzschild and Planck phases.
  Since 2002, we have seen big progress in understanding spindown and Schwarzschild phases
  by obtaining greybody factors for rotating higher dimensional black holes. Still the first and
  last phases are largely unknown.}
\label{rsbulk}
\end{figure}

\section{Progress-2: Decay}
The black hole, which is first formed in a particle collision, is expected to be highly asymmetrical.
However, in the {\bf balding phase} the black hole looses its "hair" by emitting energy and charge in the form of
gravitational radiation and gauge boson emission. At the end of the day, it will become a stationary, rotating black hole which can be
nicely described by the Myers-Perry solution ~\cite{Myers-Perry} in higher dimensions \footnote{The upper bound for the
angular momentum was estimated by taking the initial condition of the angular momentum into account using
the hoop conjecture ~\cite{Park2001}: $0 < a\equiv (n+2)J/(2 M_{bh}R_{bh})<(n+2)/2$.}. There is big room for future contribution in understanding
the balding phase. One may hope that numerical studies for dynamical black hole formation can achieve improved
understanding in the future.

\begin{figure}
    \includegraphics[width=.35\textheight]{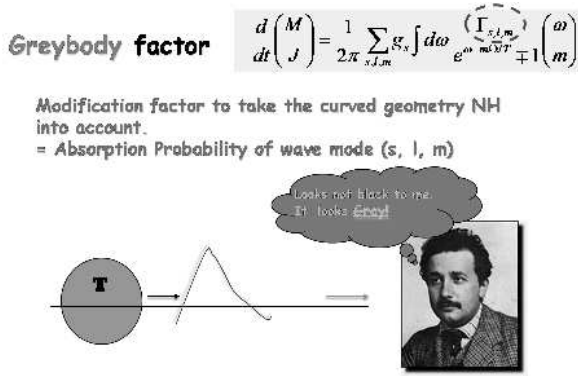}\\
  \caption{This cartoon shows the physical meaning of the greybody foctors of black holes.
 The thermal spectrum of the black hole is highly modified by the near horizon geometry and results
  in a highly anisotropic, energy dependent pattern of observed Hawking radiation by
  the observer.}
\label{greybody}
\end{figure}

The phases following the balding phase are the {\bf spindown phase} and the {\bf Schwarzschild phase}. In these phases
black hole looses its energy and angular momentum through Hawking radiation ~\cite{Hawking}.
The pattern of Hawking radiation is largely relying on the angular momentum of black hole
in spindown phase but only recent studies by Ida, Oda and myself ~\cite{IOP1,IOP2-1,IOP2-2,IOP3} and
several other authors including Kanti and her collaborators ~\cite{Kanti1,Kanti2,Kanti3}
make it possible to understand the Hawking radiation to the brane localized fields ~\cite{Emparan}. The key of recent improvement is that now {\it the
greybody factors} of the higher dimensional rotating black hole are obtained by analytic and numerical methods.
The black hole has an essentially thermal spectrum and its temperature is
  proportional to the surface gravity on the horizon. However the curved geometry in the
  vicinity of the black hole does not allow for an observer at infinity to see
  the direct thermal spectrum but only something modified from the original
  one. The greybody factor describes the modification and contains all the geometrical
  information near the event horizon. To calculate the Hawking radiation precisely,
  one has to know the greybody factors(See fig.~\ref{greybody}).
  How can we actually calculate greybody factors? Firstly we expand the `generalized Teukolsky equation' ~\cite{IOP1} in the
  vicinity of the horizon as well as at infinity. Once the solutions at each
  domain are obtained, match those two extremal solutions in the middle of parameter
  space. Then we can read out the solution in the whole domain from the event
  horizon to the infinity. Taking the ratio of ingoing and outgoing waves at
  infinity gives us the greybody factor or the absorption probability ~\cite{IOP1,IOP2-1,IOP2-2,IOP3}.

\begin{figure}
    \includegraphics[width=.55\textheight]{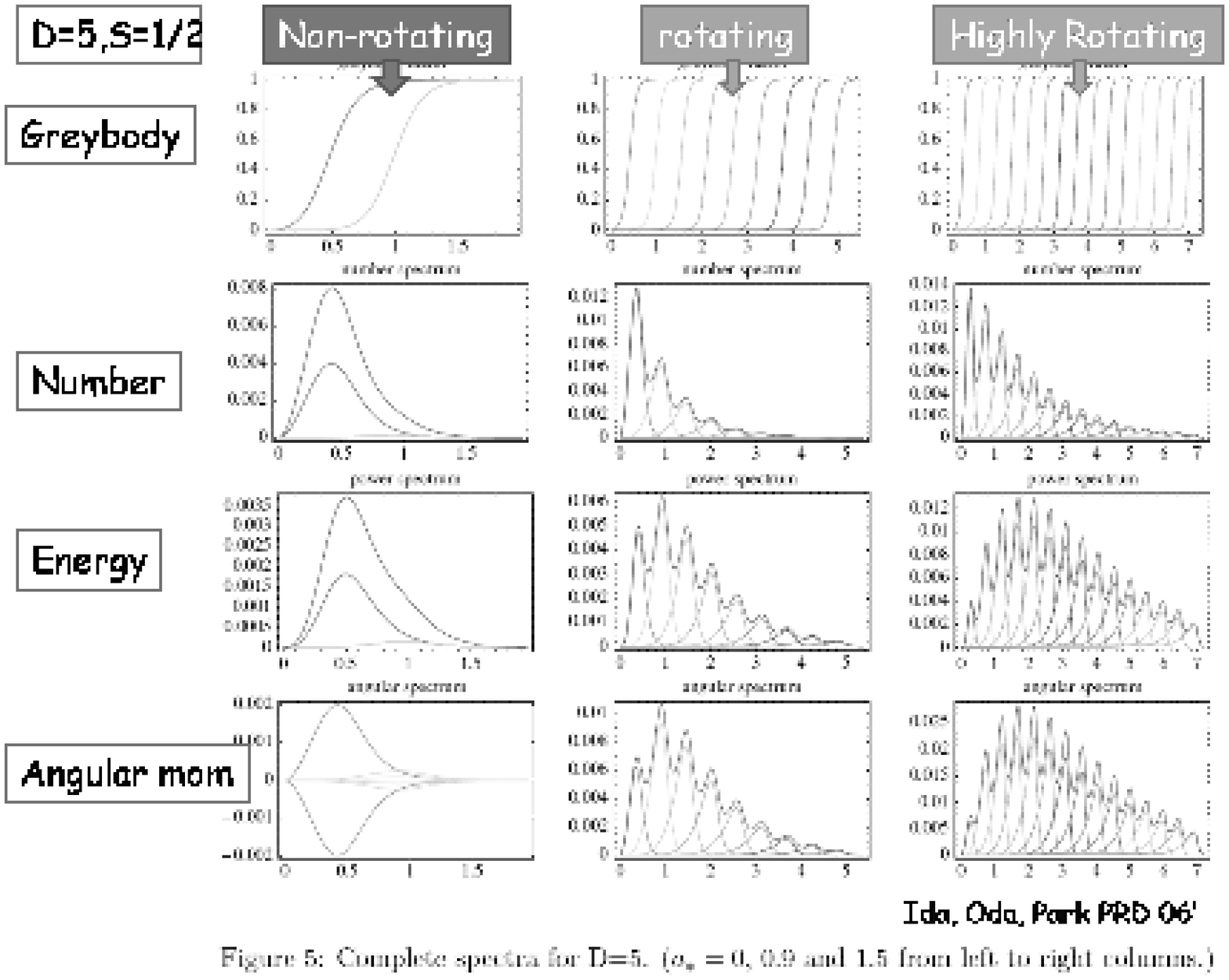}
  \caption{Here is the explicit example of obtained greybody factor, the power spectrum,
  the torque spectrum and number spectrum of five dimensional black hole to the brane
  fermion field. When black hole is non-rotating, the spectra are `simple' and `calm' but
  if it is highly rotating many angular modes with several $(l,m)$ angular quantum number
  are contributing to the spectra. That is why the task to reconstruct the mass and angular
  momentum of black hole from the detected Hawking radiation at the particle detector, in general.}
  \label{sample1}
\end{figure}

Here is a guide for future improvement.

\begin{itemize}
  \item Balding phase should be understood by dynamical simulation which can  most probably only be done
  by numerical study ~\cite{pretorius}.
  \item For $D\gg4$, highly rotating black holes, spin-2 graviton emission to the bulk can be sizable and dominant.
  Hawking radiation to spin-2 particle has been considered only for non-rotating case in higher dimensions $D>4$ ~\cite{graviton1,graviton2}.
  \item The final state of black hole, i.e. {\bf Planck Phase},  is extremely poorly understood. Full quantum gravitational consideration
  is required ~\cite{maldacena}.
  \item Details of signals should depend on many factors. (e.g. 2 to 2 dominance ~\cite{Meade}, Chromosphere ~\cite{chromo1,chromo2}, inelastic effect~\cite{inelastic}, recoil~\cite{recoil}, split-brane~\cite{rsadd, split} etc.)
\end{itemize}

\section{How many black holes will be produced}

In contrast to the original expectation ~\cite{Giddings:2001bu, Dimopoulos:2001hw}, the production cross section
of thermal black hole is not very large. First of all, for the thermal or classical black hole, the energy
threshold should be much larger than $M_D$. Let us introduce a useful parameter $x_{\rm min}$ which is defined
as the ratio of the minimum mass for the black hole and the fundamental scale :$M_{\rm min}/M_D$. Based on
entropy, $x_{\rm min}=5$ is often chosen to calculate the total cross section (see e.g. ~\cite{Giddings_review}
and references therein). With this large threshold, the cross section is largely suppressed by PDF ~\cite{Cheung}.
Another significant suppression factor is expected when we think of the case where the standard model particles
are actually bulk fields in warped extra dimensions~\cite{Rizzo:2006di}. The bulk standard model has several
nice phenomenological features and recently gained much attention (see K. Agashe's paper in this
proceedings ~\cite{agashe}). In Fig.~\ref{rsbulk} we plotted the dominant contributions to the black hole
formation by the bulk standard model field in the warped background ~\cite{park_prep}. The most dominant contribution
comes from gluon-gluon , gluon-b, gluon-longitudinal component of weak gauge boson ($W_L, Z_L$), $W_L W_L$ and $Z_L Z_L$
collisions in the range of CM energy $\sqrt{s} = 14 -100$ TeV (See Fig. ~\ref{rsbulk}). For the LHC, the total cross section is roughly as large
as $1$ fb. This number is not as huge as the first expectation but still sizable so that we will be able to observe
one hundred events with the 100 ${\rm fb}^{-1}$ luminosity which is expected for one year run. Thanks to the {\it clean}
signal of the thermal spectrum, it is plausibly expected to be easy to distinguish the black hole signals from the standard background.
For details of the signal, see the next section.

\begin{figure}
    \includegraphics[width=.35\textheight]{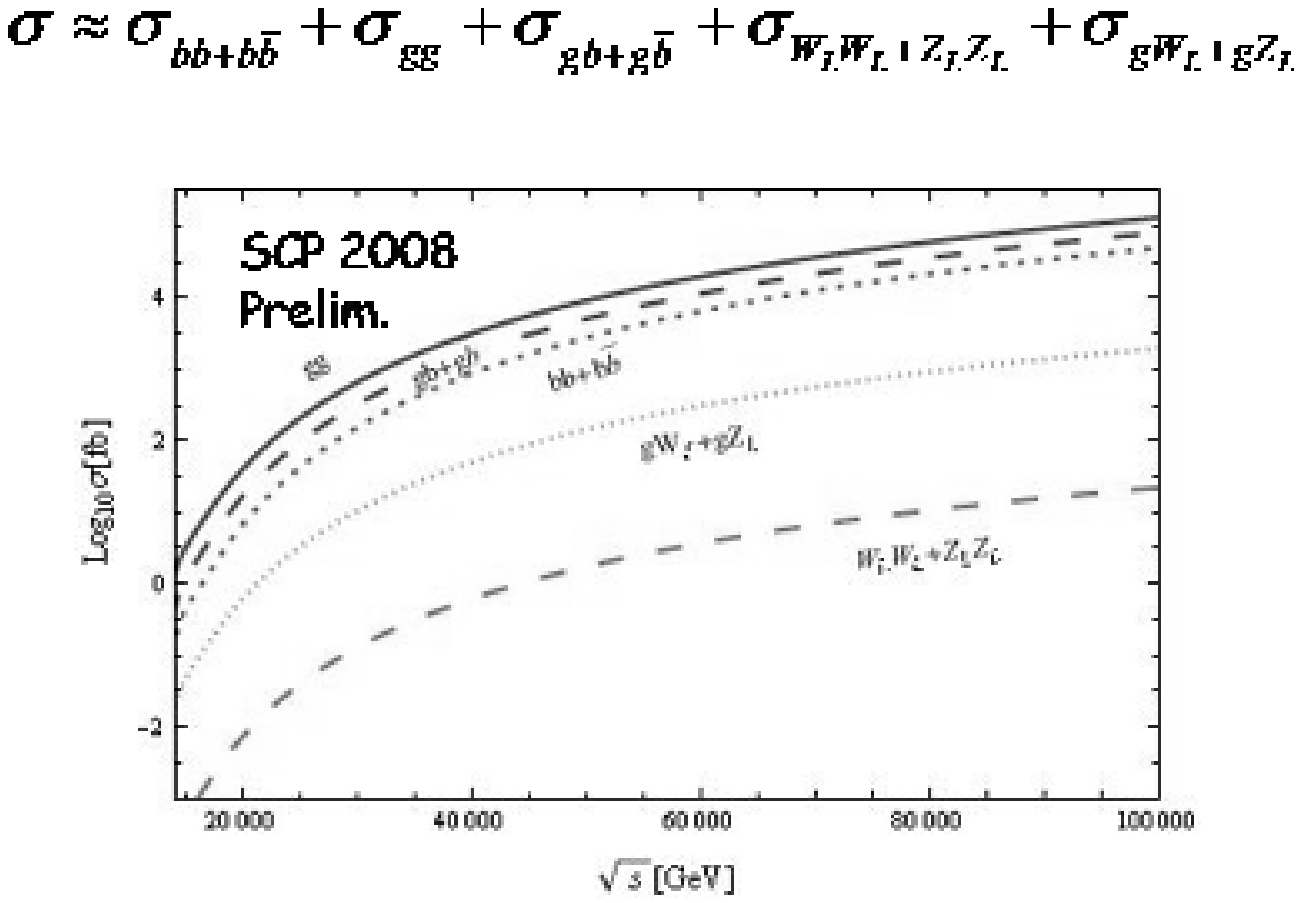}
  \caption{When the SM particles are in the bulk, most of light fermions do not
  contribute to the black hole production. The dominant contributions come from
  the fermions localized on the IR brane, the tip of gluon field which is flat
  in the bulk and the longitudinal component of $Z$ and $W$ bosons since they are
  equivalent to the Higgs.}
\label{rsbulk}
\end{figure}

\section{Black hole signal}

What is the expected signal from the black hole? All the details of the signal depends on many factors but the most robust and basic
features of black hole signals can be summarized as follows.

\begin{itemize}
  \item Large multiplicity
  \item Flavor blindness
\end{itemize}

The large multiplicity of the signal is inherited from the large entropy of the thermal black hole. Thermal black holes typically
have a large mass ($> 5 M_D$) and the Hawking radiation contains large numbers of particles or large numbers of jets, in particular.
Flavor blindness of the black hole signal can be understood since the Hawking radiation is essentially thermal. Even when we
take the greybody factors into account, flavor blindness remains. Statistically we would expect the same number of
electron, muon and tau particles in the Hawking radiation.

\begin{figure}
    \includegraphics[width=.35\textheight]{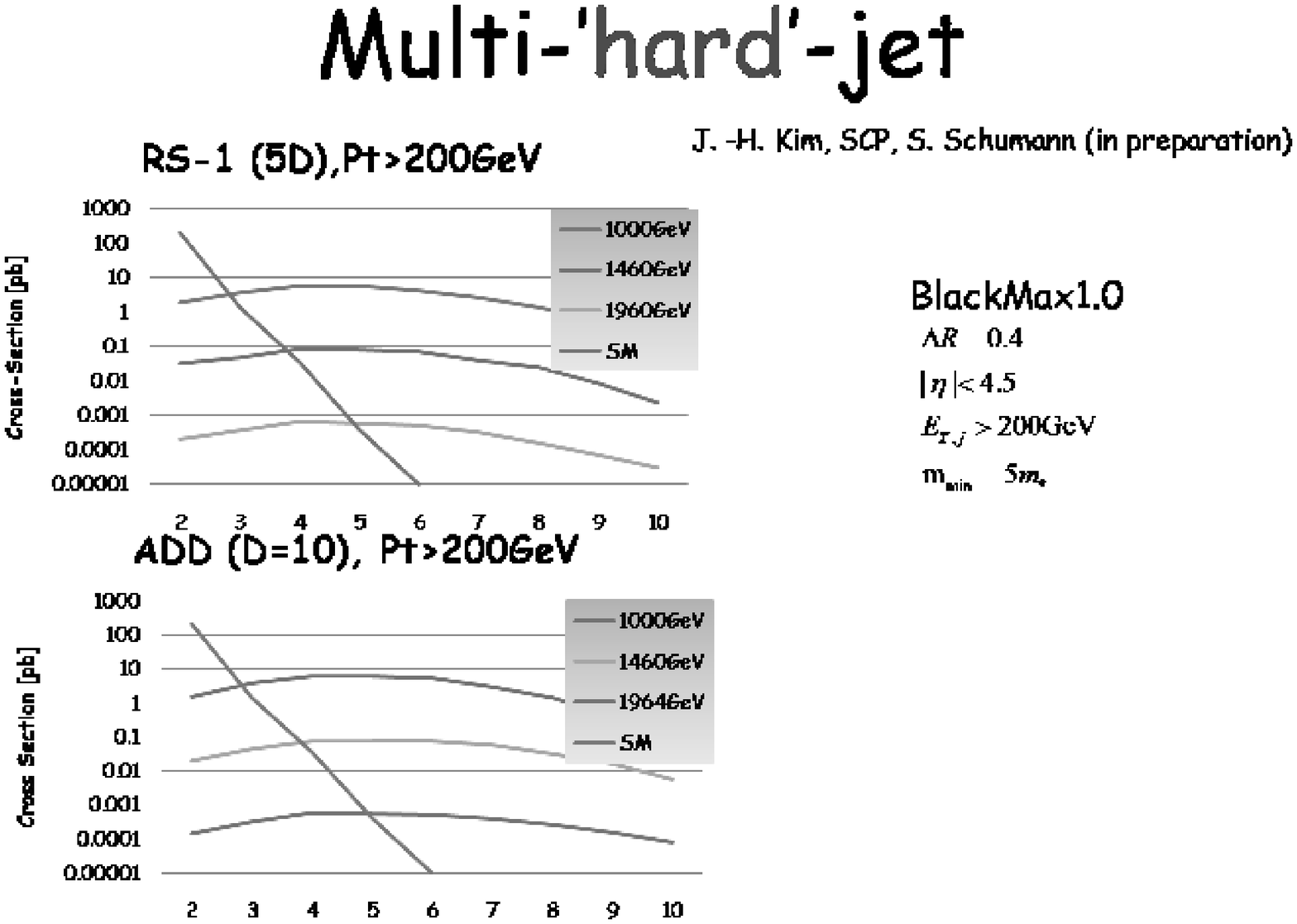}\\
    \includegraphics[width=.35\textheight]{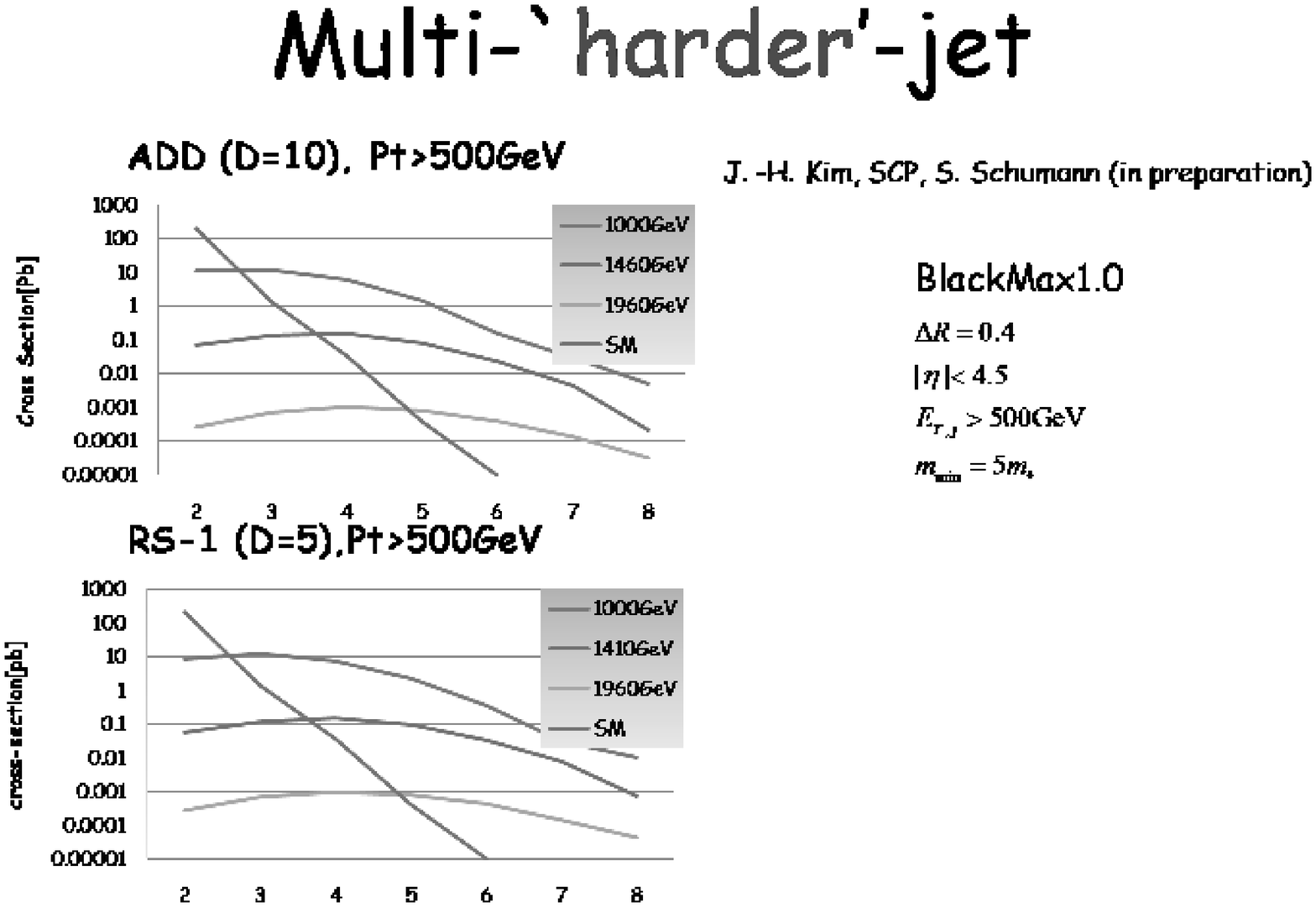}
  \caption{The number of hard jet ($P_T>200 (500) {~\rm GeV}$, $j>3$) could be a nice indication
  of observed black holes since the SM background becomes smaller very fast ~\cite{schumann}}
\label{200}
\end{figure}

For a more realistic estimation of the black hole signal, it is very helpful to have Monte-Carlo simulation code for black hole events.
Several black hole event generators have been developed ~\cite{catfish, CHARIBDIS, tanaka} to study collider signatures at the LHC but there
is only one event generator named BlackMax ~\cite{blackmax} that includes all of the black hole greybody factors up to date,
and thus can offer more realistic predictions for the LHC \footnote{James Frost (ATLAS)~\cite{Frost} informed me that the new version
of CHARIBDIS contains the relevant greybody factors as well. It is currently being tested.}.
In Fig.~\ref{200}, we plotted the number of multi-jet events ($j>3$) with the very high $P_T$ cut ($>200(500)$ GeV)
 in such a way that basically only a very small standard model background comes in. The figure clearly shows that the multi-jet
 events reduce slowly for the black hole signals but the standard model background reduces very fast as is expected by the multiple
 gauge coupling constants coming into play ~\cite{schumann}.

\section{Is the LHC safe?}

Recent analysis shows that the macroscopic effects of TeV-scale black holes should have already been seen in various
astrophysical environments if those black holes are long-lived or stable ~\cite{safety1,safety2}. This analysis
is important since there can be the extremely hypothetical but dangerous possibility that the mini black holes, which
are to be produced by the LHC, can `eat' the whole earth. If the LHC can produce black holes, there are
various places in the universe such as neutron stars and white dwarfs that similar mini black holes can
also be produced copiously by collisions of the ultra-high energy cosmic rays and nucleons in those dense objects.
Basically all those dense astrophysical objects should be in the same or a higher level of danger. Thus the lack
of any observed signals from those dense astrophysical objects ensures us that there is no risk of any significance
whatsoever from such blackholes.

\section{Summary}
If one of TeV scale gravity scenarios, such as ADD or RS models,  is correct, the LHC will have a big
chance to discover not only the existence of extra dimensions but also the deeper nature of space and time itself
by observing  the production and decay of black holes. It will provide us the first chance to check
if the current understanding of the quantum nature of black holes is right or wrong since
we will be able to know about Hawking radiation if it is there. Certainly the observation of black holes
at the particle detector will open a new era of phenomenological study of quantum gravity.

\begin{theacknowledgments}
  This talk is based on fruitful works done in collaboration with Daisuke Ida and
  Kin-ya Oda since 2002. I am grateful also to Steffen Schumann and Jihun Kim for their wonderful
  collaboration, Steve Giddings, Michelangelo Mangano, De-Chang Dai, Dejan Stojkovic, Hirotaka Yoshino, Misao Sasaki, Tetsuya Shiromizu,
  Hitoshi Murayama, Patrick Meade, Kaustubh Agashe, James Frost, Luis Anchordoqui, Alex Nielsen, Gungwon Kang and Dale Choi for discussions and cheerful communications.

\end{theacknowledgments}

\end{document}